\documentclass{aip-cp}

\newcommand{\U}[1]{\ensuremath{\mathrm{\ #1}}}
\newcommand{\UU}[2]{\ensuremath{\mathrm{\ #1^{#2}}}}

\usepackage[numbers]{natbib}
\usepackage{rotating}
\usepackage{graphicx}

\begin{document}

\title{Latest Results from VERITAS: Gamma 2016}

\author[aff1]{Jamie Holder}
\eaddress{jholder@physics.udel.edu}
\author{the VERITAS Collaboration}
\eaddress[url]{http://veritas.sao.arizona.edu}

\affil[aff1]{Department of Physics and Astronomy and the Bartol Research Institute, University of Delaware, DE 19716, USA.}

\maketitle

\begin{abstract}
The VERITAS imaging atmospheric Cherenkov telescope array has been
observing the northern TeV sky with four telescopes since summer
2007. Over 50 gamma-ray sources have been studied, including active
and starburst galaxies, pulsars and their nebulae, supernova remnants
and Galactic binary systems. We review here some of the most recent
VERITAS results, and discuss the status and prospects for
collaborative work with other gamma-ray instruments, and with
multimessenger observatories.
\end{abstract}

\section{INTRODUCTION}

\begin{figure}[h]
  \centerline{\includegraphics[width=1.0\textwidth]{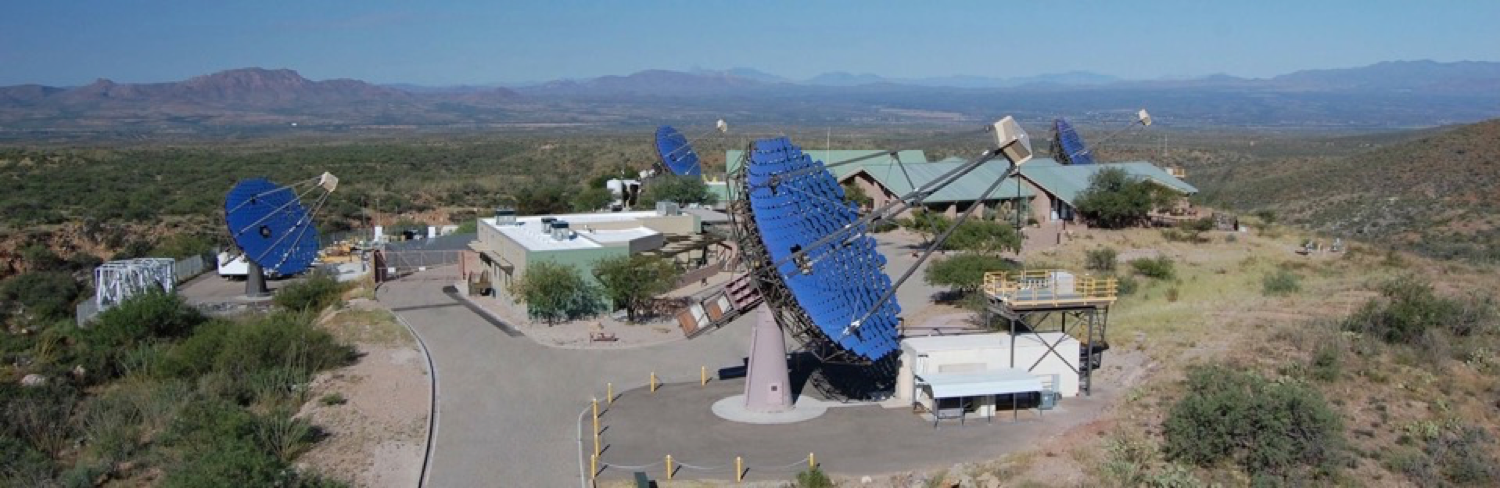}}
  \caption{The VERITAS array in its current (2016) configuration.}
  \label{VERITAS}
\end{figure}

VERITAS (Figure~\ref{VERITAS}) is an imaging atmospheric Cherenkov
telescope array, now entering its tenth year of operations. The array
consists of four identical telescopes, located at the Fred Lawrence
Whipple Observatory (FLWO) in Arizona, each with a $12\U{m}$ diameter
tessellated reflector. Cherenkov light from gamma-ray and cosmic-ray
initiated particle cascades is focused by the telescope reflectors
onto 499-pixel photomultiplier tube (PMT) cameras, which cover a
$3.5^{\circ}$ field of view.

The array has undergone two major upgrades over the past decade. The
first, in 2009, involved the relocation of the original prototype
telescope to a more favorable location, resulting in an approximate
diamond-shaped layout with sides of $\sim 100\U{m}$. This improved the
angular reconstruction capabilities, and enhanced the sensitivity of
the array. In 2012, a major overhaul of the telescope electronics saw
the installation of new trigger systems, and the replacement of all of
the photosensors with higher quantum efficiency (\textit{super
  bialkali} photocathode) PMTs. This again improved sensitivity, and
led to a dramatic enhancement of the low energy response. Thanks to
these upgrades, and to improvements in analysis and calibration tools,
VERITAS now detects a source with 1\% of the steady Crab Nebula flux
in under 25\U{hours}; less than half of the exposure required in the
original array configuration. Of similar importance is the increase in
duty cycle provided by moonlight observations. VERITAS now commonly
conducts observations with the lunar disk up to 50\% illuminated,
including with reduced PMT voltage for higher illuminations. Moonlight
observations account for as much as $\sim40\%$ of the total annual
observing yield \cite{2015ApJ...808..110A}, which averages
approximately 1300 hours per year.

\begin{figure}[h]
  \centerline{\includegraphics[width=0.6\textwidth]{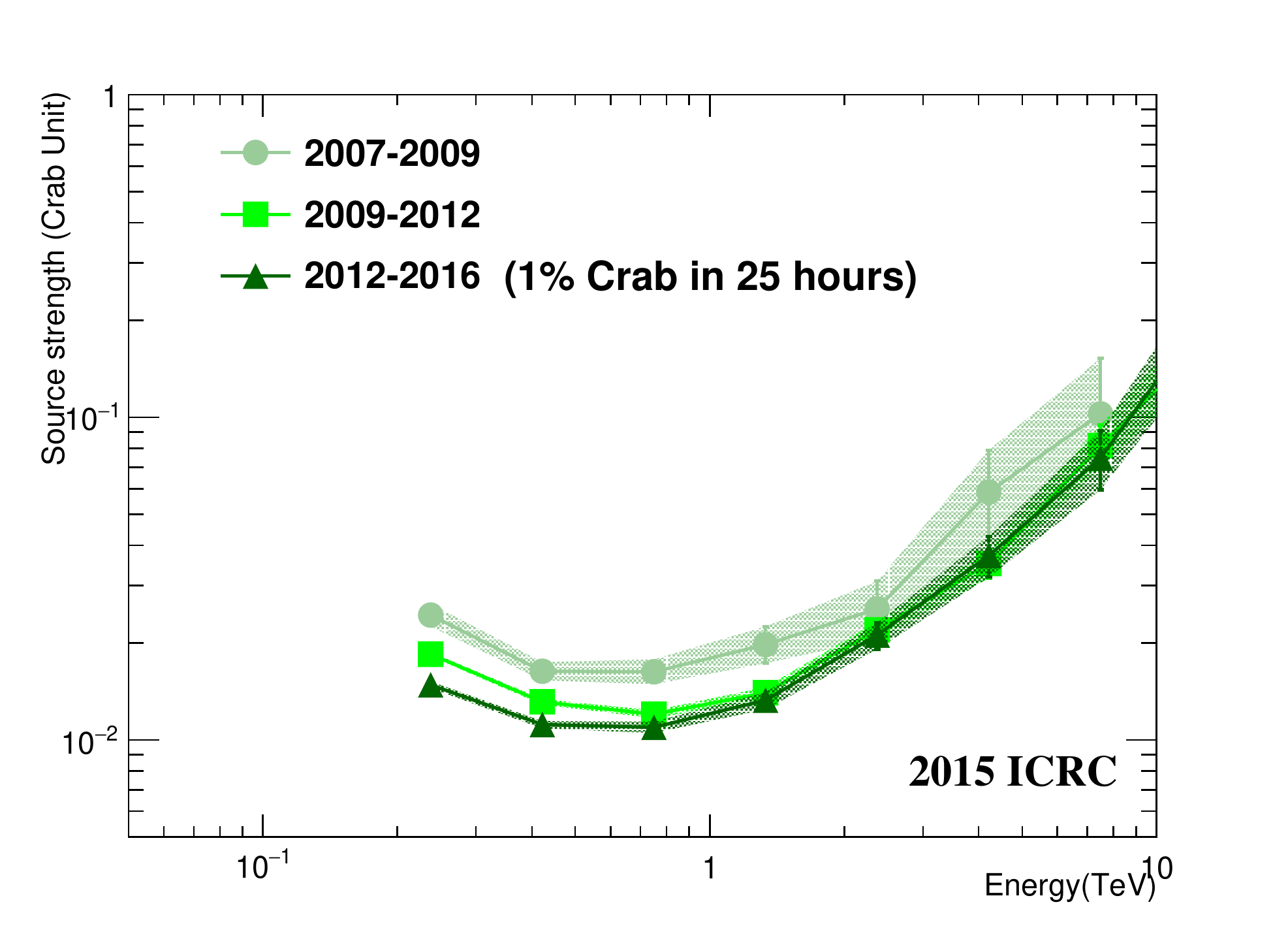}}
  \caption{VERITAS sensistivity for the three different configurations of the instrument.}
  \label{sensitivity}
\end{figure}

Figure~\ref{sensitivity} shows the differential sensitivity of VERITAS
for three different epochs. Further details are available in
\cite{2015arXiv150807070P}.

\begin{figure}[h]
  \centerline{\includegraphics[width=1.0\textwidth]{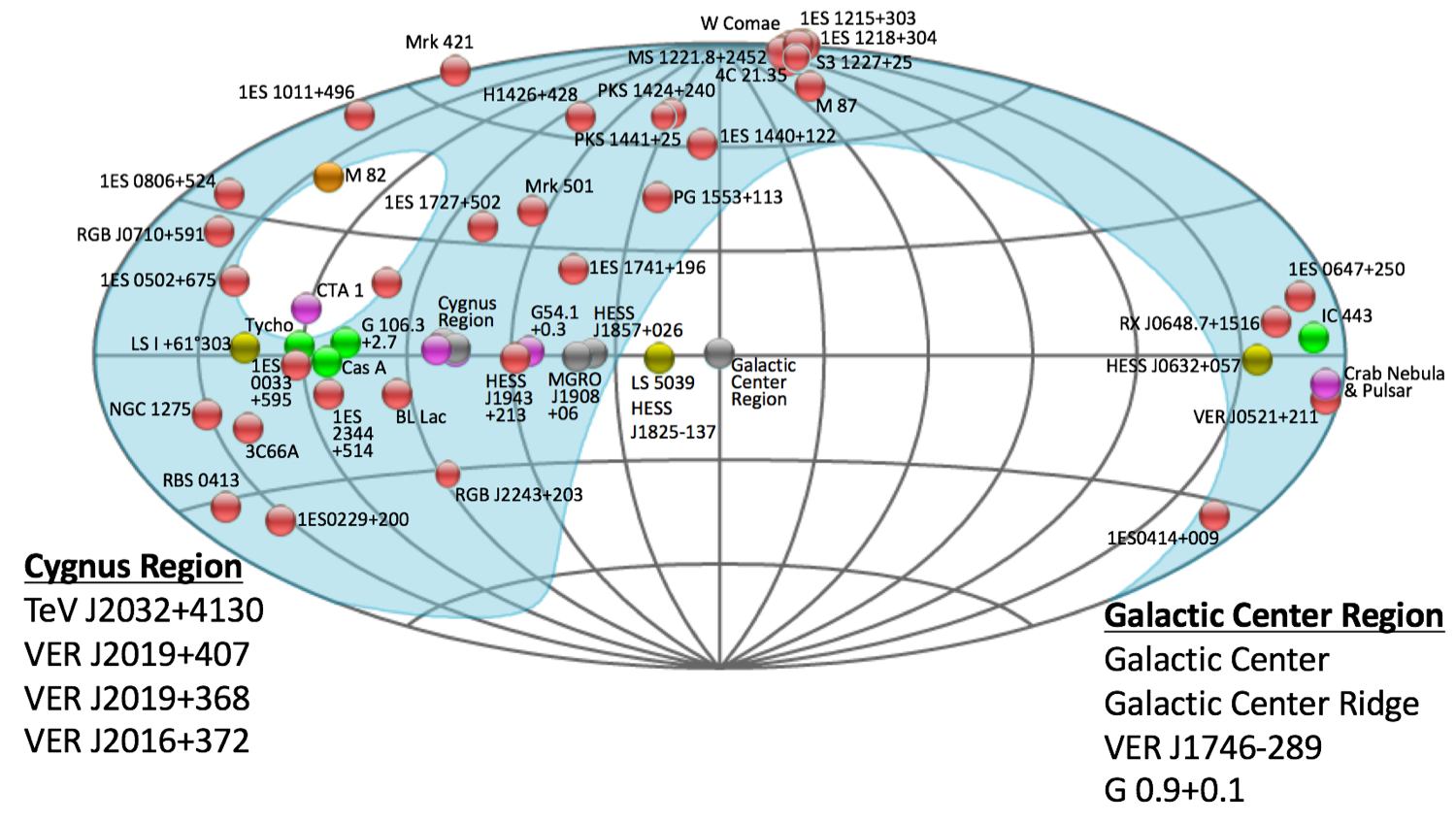}}
  \caption{The VERITAS source catalog, in Galactic coordinates, as of
    July 2016. Shaded regions indicate visibility to VERITAS above
    $55^{\circ}$ elevation. Figure modified from TeVCat ({http://tevcat.uchicago.edu}). }
  \label{catalog}
\end{figure}

The VERITAS source catalog (Figure~\ref{catalog}) now stands at 56
sources, and is composed of eight different source
classes. To make the most efficient use of limited observing time, the
balance of the observing plan has shifted from attempts to
detect new sources, to precision measurements requiring deep
exposures, or observations of variable sources during exceptional high
flux states. We summarize here some of the most interesting results
from the past two years of VERITAS operations.

\section{GALACTIC SOURCES}

\subsection{Supernova Remnants}

Supernova remnants (SNRs) are key targets for precision measurements
with VERITAS, since the details of their gamma-ray morphology and
spectra can be used to inform and constrain models of cosmic-ray
acceleration and diffusion near the source. Exposures on Tycho's SNR,
IC~443 and Cassiopeia A have all more than doubled since the original
VERITAS publications, and much of the more recent data has been taken
with the most sensitive configuration of the observatory.

\begin{figure}[h]
  \centerline{\includegraphics[width=0.52\textwidth]{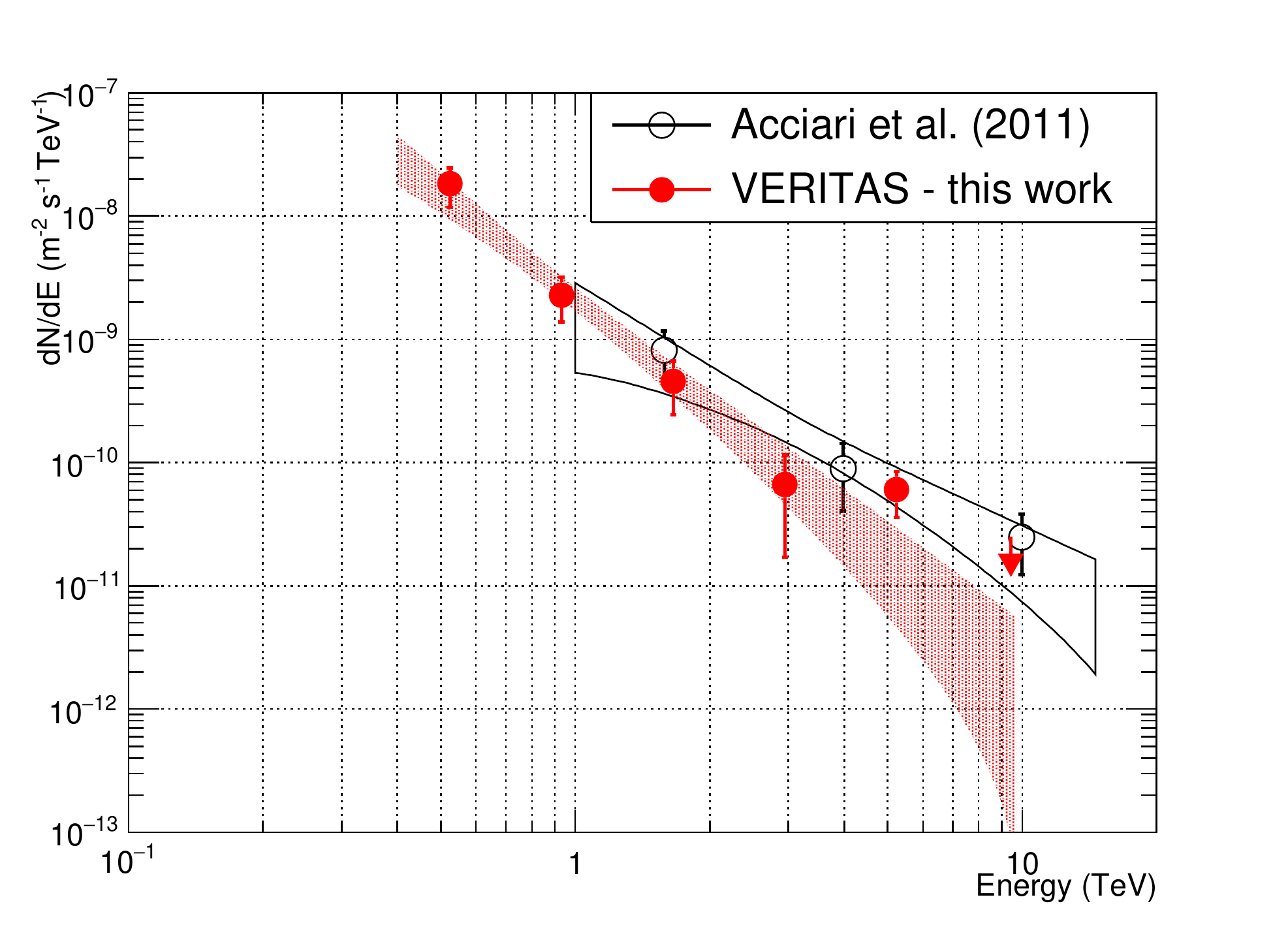}\includegraphics[width=0.48\textwidth]{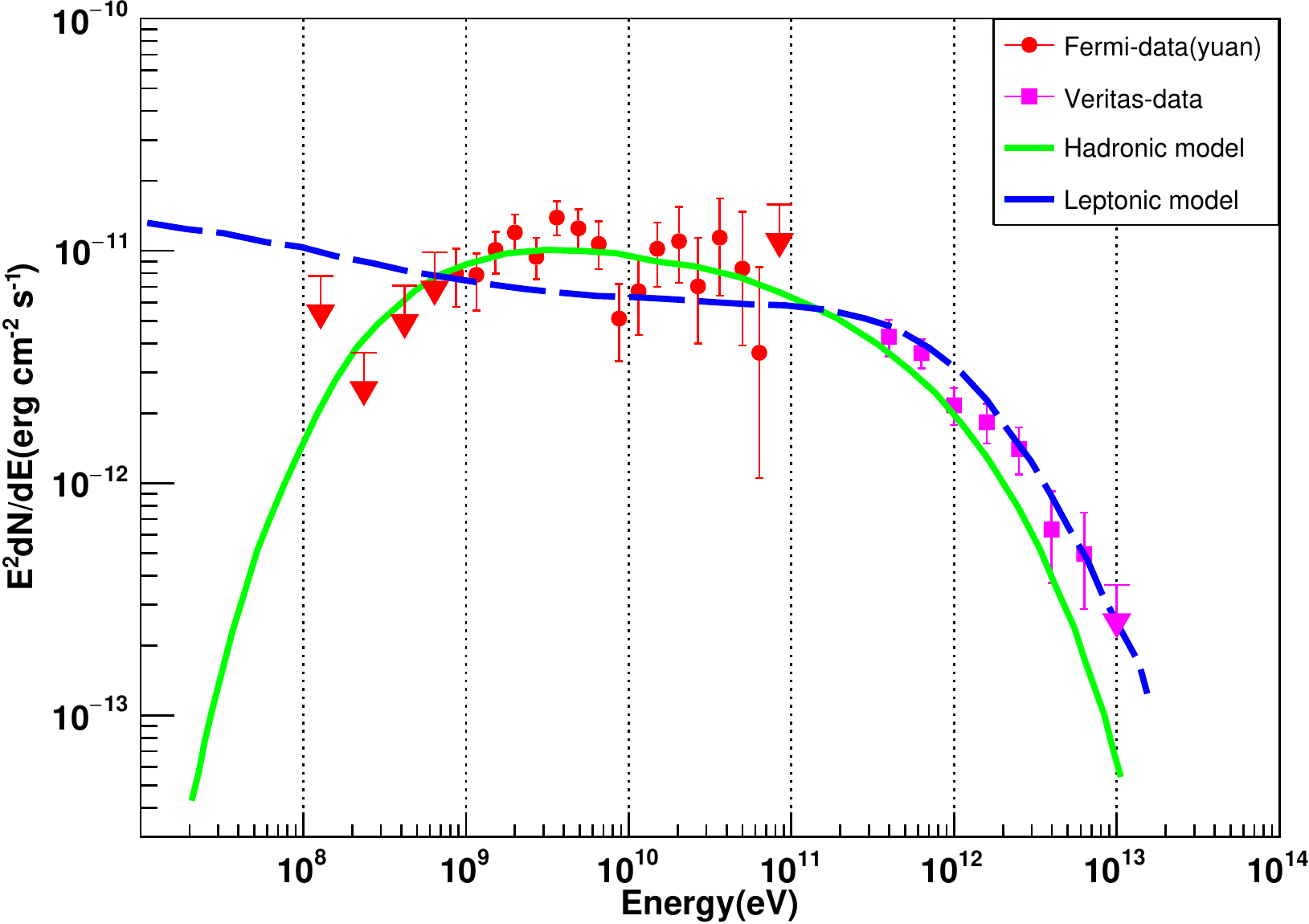}}
  \caption{VERITAS spectra of the Tycho (left) and Cas A (right) supernova remnants.}
  \label{TychoCasA}
\end{figure}

Tycho's SNR is the remnant of a historical Type Ia supernova,
well-studied over all wavelengths. The relative simplicity of the
environment around this remnant makes it a favored system for
theoretical models (e.g. \cite{2012A&A...538A..81M,
  2012ApJ...749L..26A, 2013ApJ...763...14B,
  2014ApJ...783...33S}). Gamma-ray emission from the source was
discovered by VERITAS in 2011, at the level of 0.9\% of the Crab
Nebula flux \cite{2011ApJ...730L..20A}. The broadband spectral energy
distribution (SED) has been interpreted within both hadronic and
leptonic scenarios. The updated VERITAS spectrum
\cite{2015arXiv150807068P}, shown in Figure~\ref{TychoCasA}, extends
the published measurement down in energy from $\sim800\U{GeV}$ to
$\sim400\U{GeV}$, and, while consistent, can be fit with a softer
power-law than the previous result ($\Gamma=2.92\pm0.42_{stat}$). This
may imply a steeper spectrum of accelerated particles than predicted
by standard diffusive shock acceleration models
\cite{2016A&A...589A...7M}.

Like Tycho, Cas A is a young remnant, unresolved at TeV energies, but
the relatively high gamma-ray flux allows for much more precise
spectral measurements.  Figure~\ref{TychoCasA} shows the most recent
VERITAS spectrum \cite{2015arXiv150807453K}, calculated using over 60
hours of data collected over 6 years (three times the originally
published exposure). When combined with Fermi-LAT results at lower
energies, a broken power-law fit is favored. The centroid of the
emission is consistent with an unresolved source at the the center of
the remnant, and with other measurements at TeV and GeV energies.

\begin{figure}[h]
  \centerline{\includegraphics[width=1.0\textwidth]{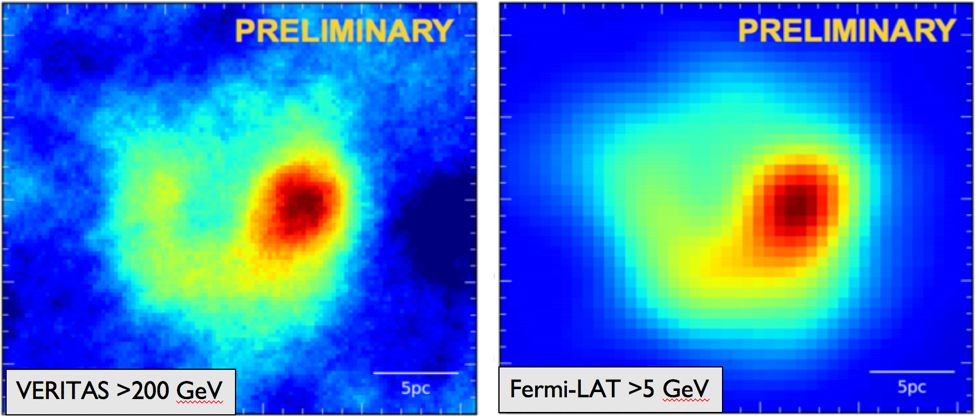}}
  \caption{VERITAS and Fermi-LAT skymaps of IC~443, showing resolved emission over the full extent of the supernova remnant.}
  \label{IC443}
\end{figure}

IC~443 is an older remnant (3-30\U{kyr}) expanding into an extremely
inhomogeneous environment. It has been established as a
hadronic accelerator, through the measurement of a pion bump signature
in the MeV-GeV range \cite{2013Sci...339..807A}. The initial TeV
gamma-ray detections \cite{2007ApJ...664L..87A, 2009ApJ...698L.133A}
identified a single site of emission co-located with dense molecular
clouds, with which the SNR is interacting. Updated results from
VERITAS now resolve emission from the entire shell of the remnant, as
illustrated in Figure~\ref{IC443} \cite{2015arXiv151201911H}. Also
shown is the $>1\U{GeV}$ map from Fermi-LAT, which correlates well
with the TeV emission \cite{2016AAS...22723810H}. The spectrum
steepens from the GeV to the TeV regime (from $\sim2.3$ to $\sim2.9$),
and this behaviour is remarkably uniform across the entire remnant,
despite the widely varying environmental conditions and integrated
flux values. The interpretation of these results is ongoing, but
IC~443 clearly provides a valuable laboratory for the exploration of
particle acceleration and diffusion over a wide range of environmental
conditions in a single object.

\subsection{Gamma-ray Binary Systems}

\begin{figure}[h]
  \centerline{\includegraphics[width=0.45\textwidth]{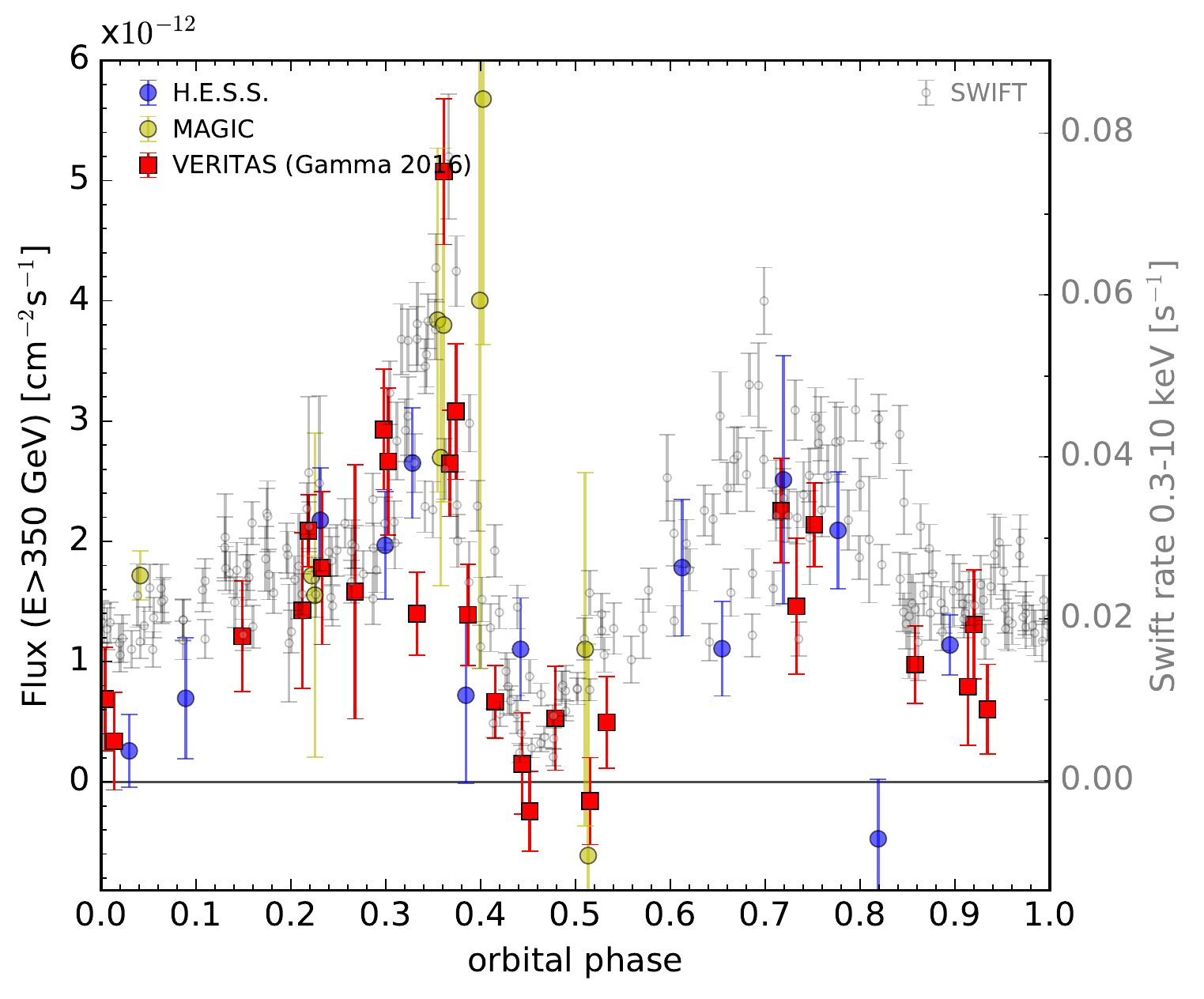}{\includegraphics[width=0.55\textwidth]{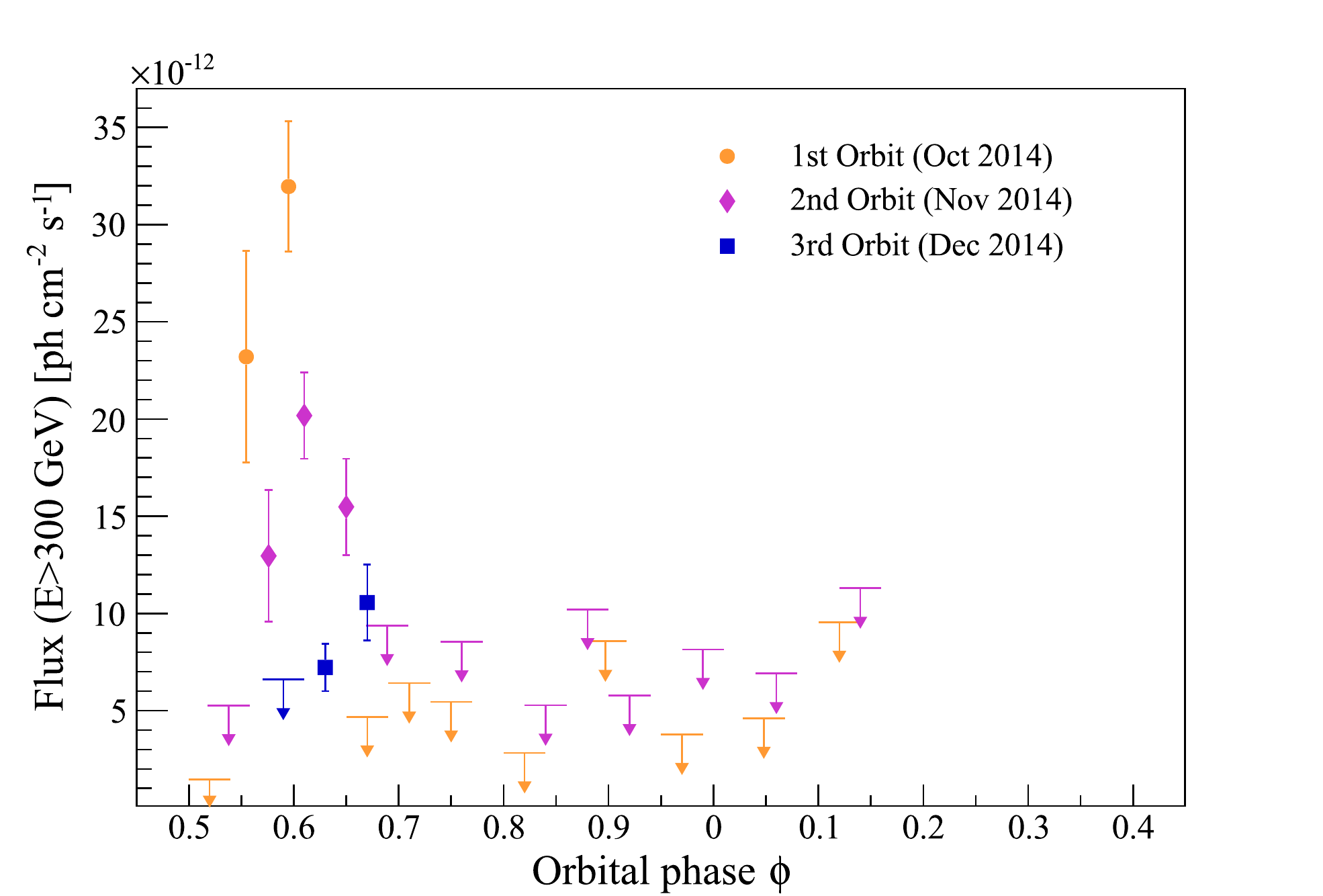}}}
  \caption{VERITAS observations of HESS~J0632+057 (left) and bright flaring activity from LS~I$+61^{\circ}303$ in 2014, reaching 30\% of the Crab Nebula flux (right).}
  \label{Binaries}
\end{figure}

Precision measurements of gamma-ray binary systems with VERITAS have
also been conducted over the almost decade-long baseline. These attempt to
provide dense and complete sampling of the gamma-ray lightcurve over
all orbital phases, as well as to monitor the systems for unusual flux
states. New results on HESS~J0632+057 are presented at this conference
\cite{HESSJ0632here}. The full 315-day orbit is now sampled by VERITAS
observations, providing clear detections and spectral measurements
over all phases in which emission is detected
(Figure~\ref{Binaries}). Also shown are observations of
LS~I$+61^{\circ}303$ which, in 2014, revealed exceptionally bright TeV
flares, reaching $\sim30\%$ of the Crab Nebula with variability on
daily timescales \cite{2016ApJ...817L...7A}. Plans for the coming
seasons include monitoring of PSR~J2032+4127 as it approaches
periastron in 2017-2018. This system is a recently identified 20-30
year-period pulsar-Be star binary co-located, and possibly associated,
with TeV 2032+4130 \cite{2015MNRAS.451..581L}.

\subsection{The Galactic Center Region}
\begin{figure}[h]
  \centerline{\includegraphics[width=0.8\textwidth]{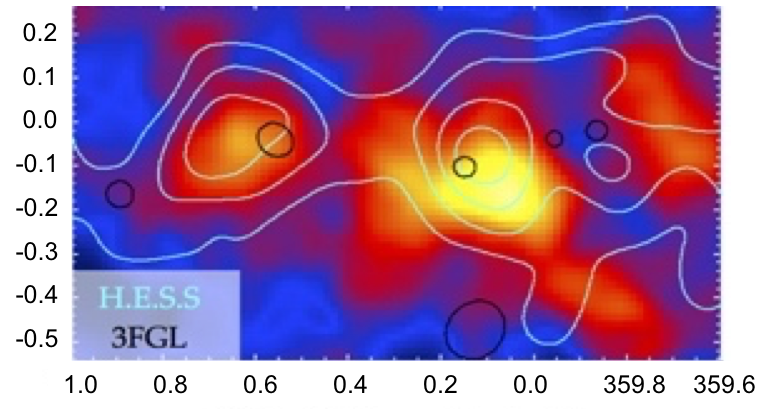}}
  \caption{VERITAS map of the Galactic Center region (in Galactic coordinates), with unresolved emission from the direction of SgrA* and SNR G0.9+0.1 subtracted. Emission contours from H.E.S.S. are also shown (cyan), along with error ellipses for Fermi-LAT sources (black). See \cite{2016ApJ...821..129A} for details.}
  \label{GC}
\end{figure}

VERITAS results on a deep exposure ($85\U{hours}$) of the Galactic
Center region were also recently published
\cite{2016ApJ...821..129A}. With an average elevation angle of
$\sim30^{\circ}$, the energy threshold for this analysis is
$2\U{TeV}$, complementing studies by H.E.S.S. Point-like emission from
the direction of both SgrA* and the composite SNR G0.9+0.1 has been
measured. When these sources are subtracted from the excess maps,
residual emission is observed along the Galactic Ridge, including a
relatively bright component close to SgrA*, which we label
VER~J1746-289 (Figure~\ref{GC}).

\section{EXTRAGALACTIC SOURCES}

With its location in the northern hemisphere, much of VERITAS science
naturally focuses on the extragalactic sky, and active galactic nuclei
(AGN) are the most numerous source class in the catalog. VERITAS has
detected 34 AGN, the majority of which are high- or
intermediate-frequency peaked BL Lacertae objects (HBLs/IBLs). The
remainder are the flat spectrum radio quasars (FSRQs), PKS 1222+216
and PKS 1441+25, the nearby radio galaxies M87 and NGC 1275, and the
starburst galaxy M82. Multiwavelength monitoring programs provide
contemporaneous measurements of the broadband SEDs of many of these
sources, in particular through close collaboration with the Swift
X-Ray Telescope team. VERITAS blazar results are covered in more
detail elsewhere in these proceedings \cite{Blazarshere,
  HESSJ1943here}, as is their use as probes of the extragalactic
background light (EBL) and intergalactic magnetic fields
\cite{EBLIGMFhere}. Here we highlight just a couple of the most
exciting recent extragalactic source results.

\subsection{PKS 1441+25}
\begin{figure}[h]
  \centerline{\includegraphics[width=1.0\textwidth]{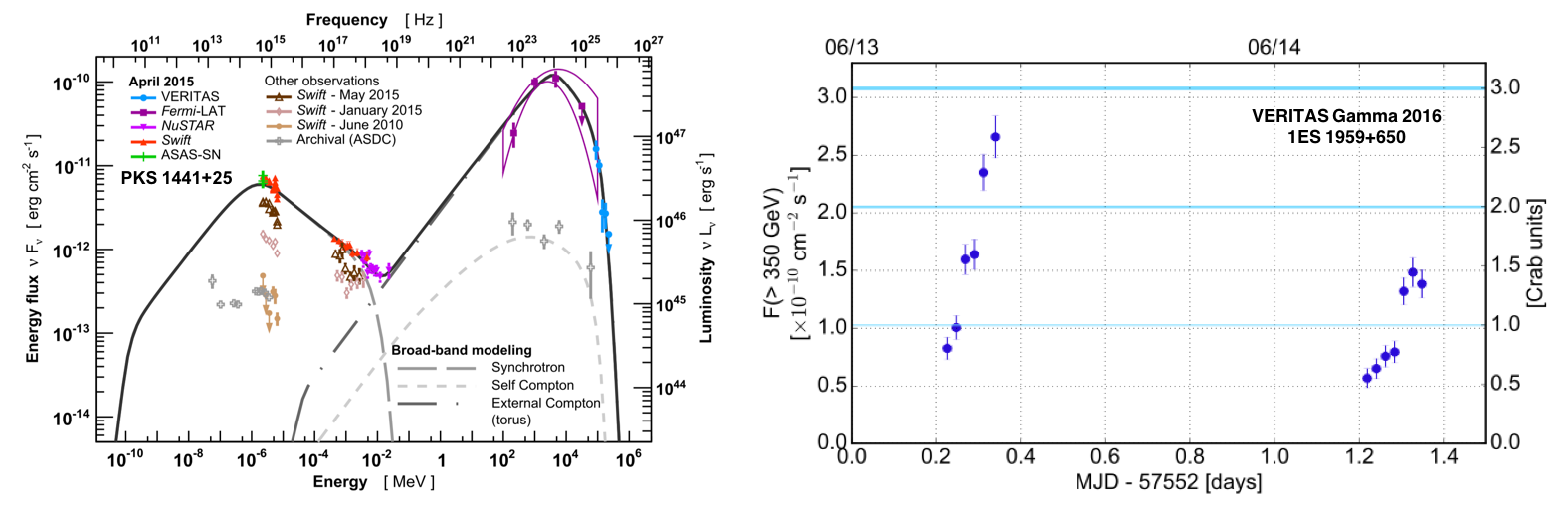}}
  \caption{The broad-band SED of the distant FSRQ, PKS 1441+25 (left),
    and (right) a portion of the VERITAS lightcurve for 1ES~1959+650 during an active period
    in June 2016.}
  \label{blazars}
\end{figure}

PKS~1441+25 is one of only two FSRQs detected by VERITAS
\cite{2015ApJ...815L..22A} and, at a redshift of $z=0.939$, is the
second-most distant source in the very high energy sky. Observations
in April 2015, triggered by alerts from Fermi-LAT
\cite{2015ATel.7402....1P} and MAGIC \cite{2015ATel.7416....1M},
allowed a precise measurement of the emission during a high state, in
which the flux reached $\sim5\%$ of the Crab Nebula flux above
$85\U{GeV}$. The spectral index, $\Gamma=-5.4\pm0.5$, is the softest
yet measured by VERITAS for any source, corresponding to an intrinsic
index of $\Gamma=-3.4\pm0.5$ after correction for absorption by the
EBL (Figure~\ref{blazars}). No emission is detected above
$\sim200\U{GeV}$, illustrating the importance of the energy threshold
reduction provided by the 2012 camera and trigger upgrades for
extragalactic science. Gamma-ray opacity arguments, as well as
multiwavelength correlations, indicate that the origin of the emission
must lie far from the center of the AGN, outside of the broad line
region. The extreme distance of the source, coupled with the highly
significant measurement of its spectrum, allow to place constraints on
the intensity of the near-ultraviolet to near-infrared EBL which are
comparable to those derived from stacking multiple sources with lower
redshifts, and agree with EBL estimates from galaxy counts.

\subsection{1ES 1959+650}

1ES~1959+650 was among the first TeV blazars to be detected
\cite{2003ApJ...583L...9H} and holds particular interest as the
clearest example of a TeV source which has exhibited ``orphan''
flaring - a prominent gamma-ray flare with no X-ray counterpart
\cite{2004ApJ...601..151K}. The source has been in an extended active
state since summer 2015, and VERITAS observed flares at the level of
the Crab Nebula flux in fall 2015 \cite{2015ATel.8148....1M}. Further
flaring activity in 2016 triggered additional VERITAS monitoring,
revealing the source to be in an unprecedented high state, with flares
exceeding 2.5 Crab (Figure~\ref{blazars}). Cooperation and
communication between the northern hemisphere gamma-ray instruments,
Fermi-LAT and Swift places this among the best-sampled extreme blazar
flaring events, with observations often conducted multiple times
within a 24-hour period.

\section{OTHER TOPICS}
Gamma-ray astronomy is an important and growing field, but imaging
atmospheric Cherenkov telescopes (IACTs) provide the means to study
more than gamma-ray emission from specific astrophysical objects. We
summarize some recent updates from VERITAS below.

\subsection{Dark Matter Searches}

The most recent update on the search for gamma rays from
WIMP annihilation in dwarf spheroidal galaxies with VERITAS was
presented at the 2015 ICRC \cite{2015arXiv150901105Z}. As the exposure
for targets with the highest estimated luminosity due to dark matter
annihilation approaches $\sim100\U{hours}$, further deep observations
of individual targets provide diminishing returns, and are prone to
systematic biases. An alternative approach, which offers considerable
improvements, is to combine the data from multiple sources, weighting
individual events according to their energies, locations, and the
properties of the instrumental and astrophysical backgrounds
\cite{2015PhRvD..91h3535G}. Using this method, the combination of over
200 hours of observations, spread over 4 dwarf spheroidal galaxies,
has been used to provide improved limits to the annihilation
cross-section.

\subsection{Cosmic-Ray Electrons}

\begin{figure}[h]
  \centerline{\includegraphics[width=0.6\textwidth]{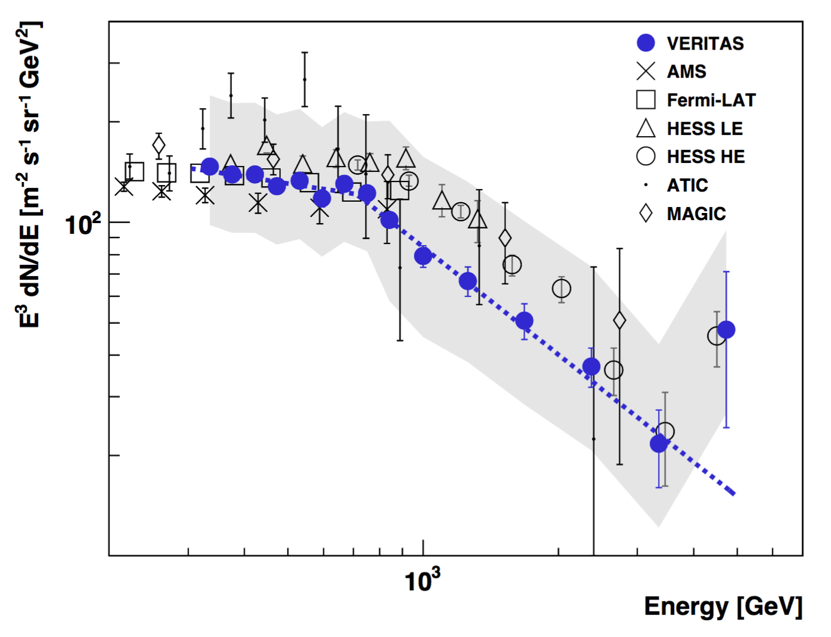}}
  \caption{VERITAS cosmic-ray electron spectrum (blue circles), fit with two power-laws with a break energy of $710\pm40\U{GeV}$. The gray band shows the systematic uncertainty on the VERITAS measurement \cite{2015arXiv150806597S}.}
  \label{physics}
\end{figure}

The cosmic-ray electron/positron spectrum provides a direct
measurement of cosmic-ray acceleration and diffusion in our local
Galactic neighbourhood. Interest in this area has grown with the
observation, by multiple instruments, of a positron component which
increases up to an energy of $\sim200\U{GeV}$. While they are unable
to easily discriminate electrons and positrons, the large effective
collection area of IACTs allows to extend measurements of the spectrum
of cosmic-ray electrons beyond those energies probed by AMS and
Fermi-LAT. VERITAS has now measured the combined electron-positron
spectrum up to a few TeV \cite{2015arXiv150806597S}. The results are consistent with those from
H.E.S.S., and the spectrum is best fit with two power-laws with a
break energy of $710\pm40\U{GeV}$ (Figure~\ref{physics}).

\subsection{Optical Transients}
With their large mirror area, IACTs are powerful optical instruments,
and probe a unique temporal parameter space. VERITAS has recently
developed an efficient method to search for ultra-fast optical
transients from astronomical objects, sensitive to nanosecond pulses
with fluxes as low as $\sim1\U{photon}\UU{m}{-2}$. We have applied
this technique to search for optical pulses in serendipitous archival
observations of KIC~8462852, a star with a peculiar variability
pattern which has been suggested as an exceptional target for SETI
searches \cite{2016ApJ...818L..33A}. Unsurprisingly, no emission was
detected, but the technique is now established, and can be applied to
searches of the extensive VERITAS data archives for brief optical
transient events.

\section{NEW PARTNERS}

VERITAS has developed ongoing partnerships with many collaborations
around the world, working both in gamma-rays and at other wavelengths,
as well as with multi-messenger astroparticle physics
observatories. Collaborative efforts with two relatively new
facilities have recently enhanced the scientific program of VERITAS.

\subsection{VERITAS and IceCube}

\begin{figure}[h]
  \centerline{\includegraphics[width=1.0\textwidth]{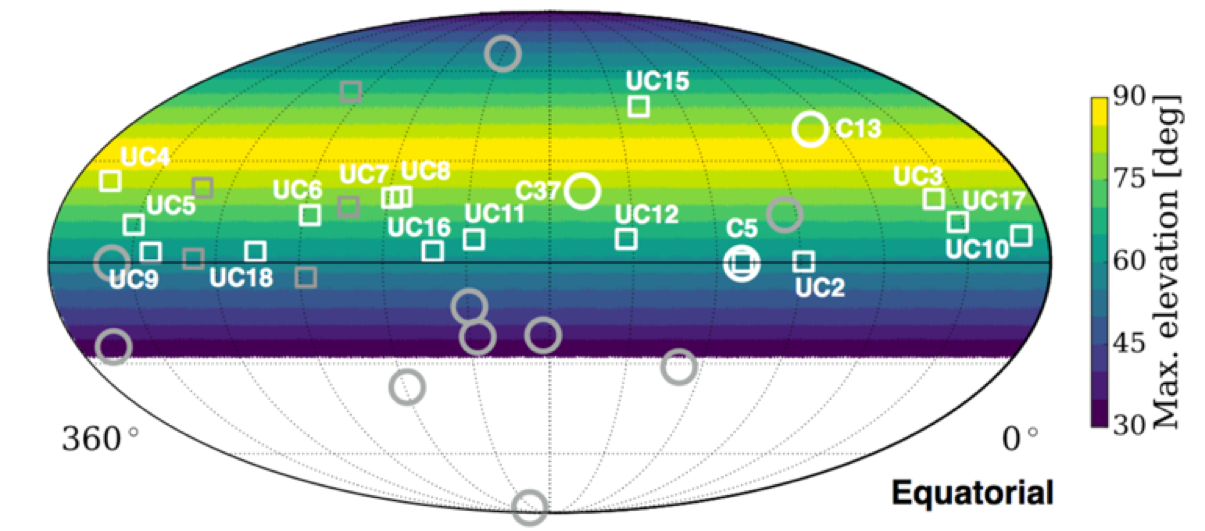}}
  \caption{IceCube neutrino locations observed by VERITAS. The labeled
    events all have relatively small angular uncertainty (muon track
    events, with $\sim1^{\circ}$ angular resolution) and a high
    probability ($>40\%$) of astrophysical
    origin \cite{2015arXiv150900517S}.}
  \label{IceCube}
\end{figure}

The IceCube observatory has detected an isotropic flux of TeV-PeV
astrophysical neutrinos, whose origin remains unclear
\cite{2014PhRvL.113j1101A}. IACT arrays, including VERITAS, can
attempt to identify a gamma-ray counterpart, or to constrain the
properties of the neutrino source population. VERITAS has made
follow-up observations of 18 well-located IceCube events; those
containing muon tracks, with $<1^{\circ}$ angular uncertainty
(Figure~\ref{IceCube}). No significant emission has been detected, and
upper limits are typically a few \% of the Crab Nebula flux
\cite{2015arXiv150900517S}.

The sources of astrophysical neutrinos may be variable, or have a
variable component. In this case, prompt follow-up observations of any
neutrino event could be the key to the discovery of a gamma-ray
counterpart. VERITAS and MAGIC have both taken part in a long-running
program, responding to potential neutrino flares originating from the
direction of a list of candidate sources (mostly blazars)
\cite{2015PoS1052}. With the identification of a high energy
astrophysical neutrino flux, more stringent conditions can now be
applied to ensure that follow-up observations are targeting the
locations of astrophysical neutrinos (as opposed to fluctuations in
the background of atmospheric neutrinos). Using infrastructure
developed for gamma-ray burst studies, IceCube is now providing prompt
Gamma-ray Coordinates Network (GCN) alerts for high-energy starting
events (HESE) through the astrophysical multimessenger observatory
network (AMON). VERITAS received its first alert in April 2016, with
observations starting after 193 seconds.  Almost all of the data were
taken under bright moonlight conditions. Unfortunately, a revised
location, issued $18\U{hours}$ after the initial alert, placed the
neutrino point of origin outside of the field of view of the VERITAS
observations. The revised location was targeted by VERITAS the
following night, but showed no excess. VERITAS flux upper limits were
posted as a GCN circular. The prompt response provided a useful
demonstration of the capabilities of the alert program, and
observations will continue over the coming observing seasons.

\subsection{VERITAS and HAWC}
The most recent partner facility to be fully commissioned is the High
Altitude Water Cerenkov (HAWC) observatory, which observes the
northern hemisphere very high energy gamma-ray sky continuously from
its location in Mexico. HAWC is able to provide alerts to the TeV
observatories, which allows prompt follow-up of extreme flaring events
(e.g. \cite{2016ATel.9137....1B}). The much higher instantaneous
sensitivity of IACTs can then be used to study such events in detail
and VERITAS, located at almost the same longitude as HAWC, is
particularly well-suited to this task.

\begin{figure}[h]
  \centerline{\includegraphics[width=1.0\textwidth]{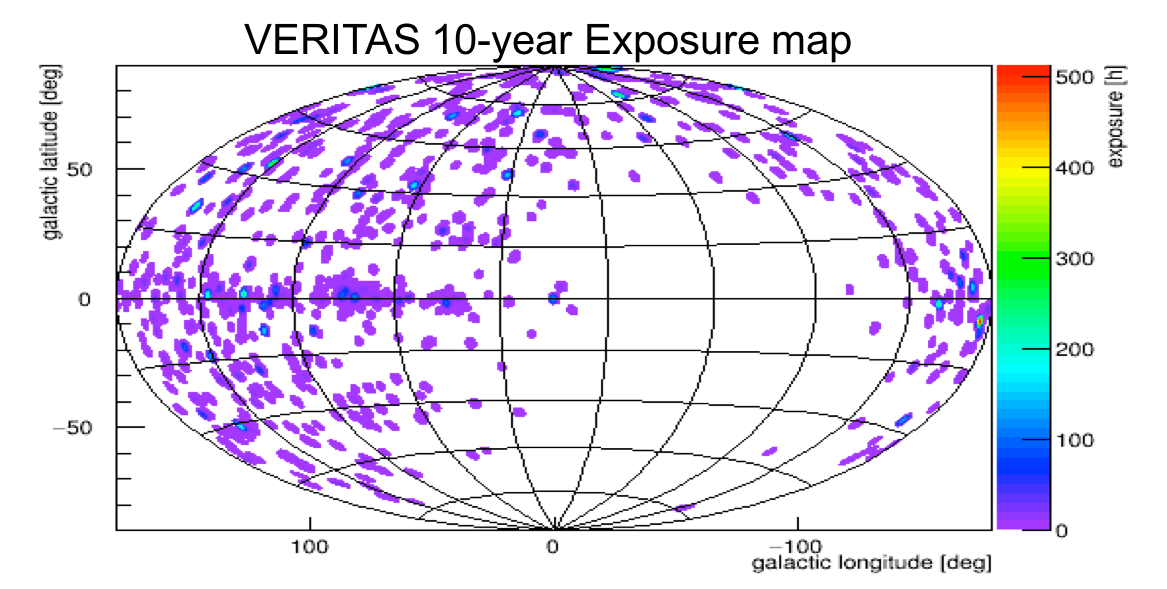}}
  \caption{The VERITAS exposure map, after 10 years of observations.}
  \label{exposure}
\end{figure}

The wide-field survey capabilities of HAWC provide another promising
avenue to new science. HAWC has recently released a preliminary skymap
based on 341 day's live-time with almost the full detector in
operation. Further results were shown in detail at this meeting
\cite{HAWC1here, HAWC2here}. While the VERITAS field-of-view
($\Phi=3.5^{\circ}$) is much smaller than HAWC, VERITAS has been in
operation since 2007, and has archival observations of a significant
fraction of the celestial sphere (Figure~\ref{exposure}). This
includes exposures, totalling $100\U{hours}$, which cover at least 10
of the sources and source candidates in the preliminary HAWC catalog
which do not have an existing TeV counterpart. A consistent
re-analysis of these data using the most recent analysis tools is in
progress, and the results will be presented at future meetings. We
note that there exists some overlap with the VERITAS survey of the
Cygnus region, an update to which was presented here
\cite{Krausehere}.

Dedicated follow-up observations have been made of two particularly
interesting regions identified by HAWC. The first is the region close
to SNR~G54.1+0.3/PSR~1930+1852. VERITAS detected a gamma-ray
counterpart to this object, VER~J1930+188, in 2010
\cite{2010ApJ...719L..69A}. The emission is point-like, with a
spectral index of $\Gamma=-2.39\pm0.23_{stat}\pm0.30_{sys}$, and is
likely associated with the pulsar wind nebula. The HAWC view of this
region displays emission from the location of the VERITAS source
(2HWC~J1930+188), along with two further sources located $0.8^{\circ}$
(2HWC~J1927+187) and $1.2^{\circ}$ (2HWC~J1928+178) from
PSR~1930+1852. Both of these excesses are close to relatively high
spin-down luminosity pulsars (PSR~J1928+1746 and PSR~B1925+18.8),
which may provide plausible candidates for the HAWC emission. We note
that the original VERITAS publication already placed limits to
point-like emission from PSR~J1928+1746 at the level of
$F(>1\U{TeV})<2.6\times10^{-13}\UU{cm}{-2}\UU{s}{-1}$.

\begin{figure}[h]
  \centerline{\includegraphics[width=0.48\textwidth]{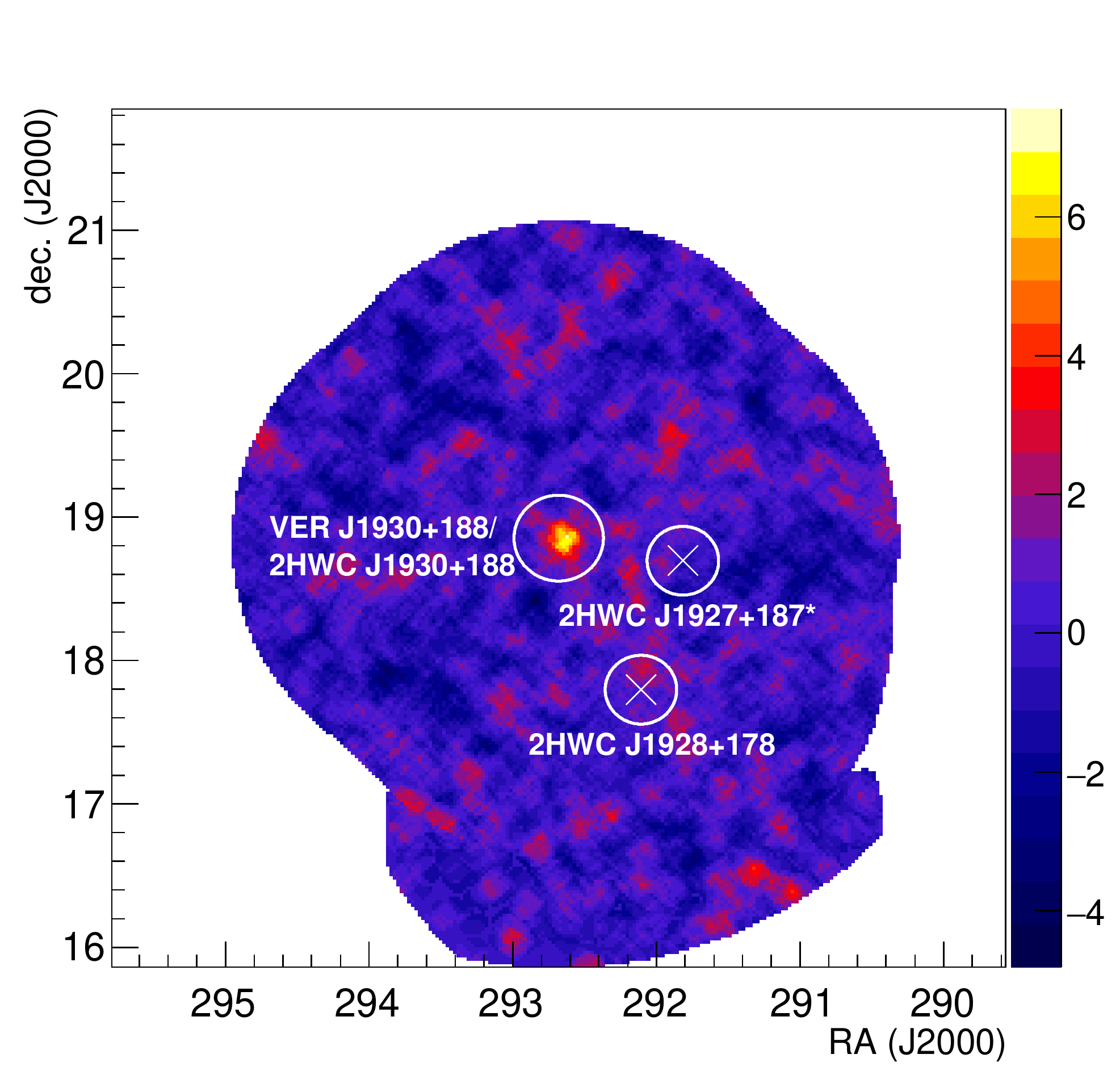}\includegraphics[width=0.52\textwidth]{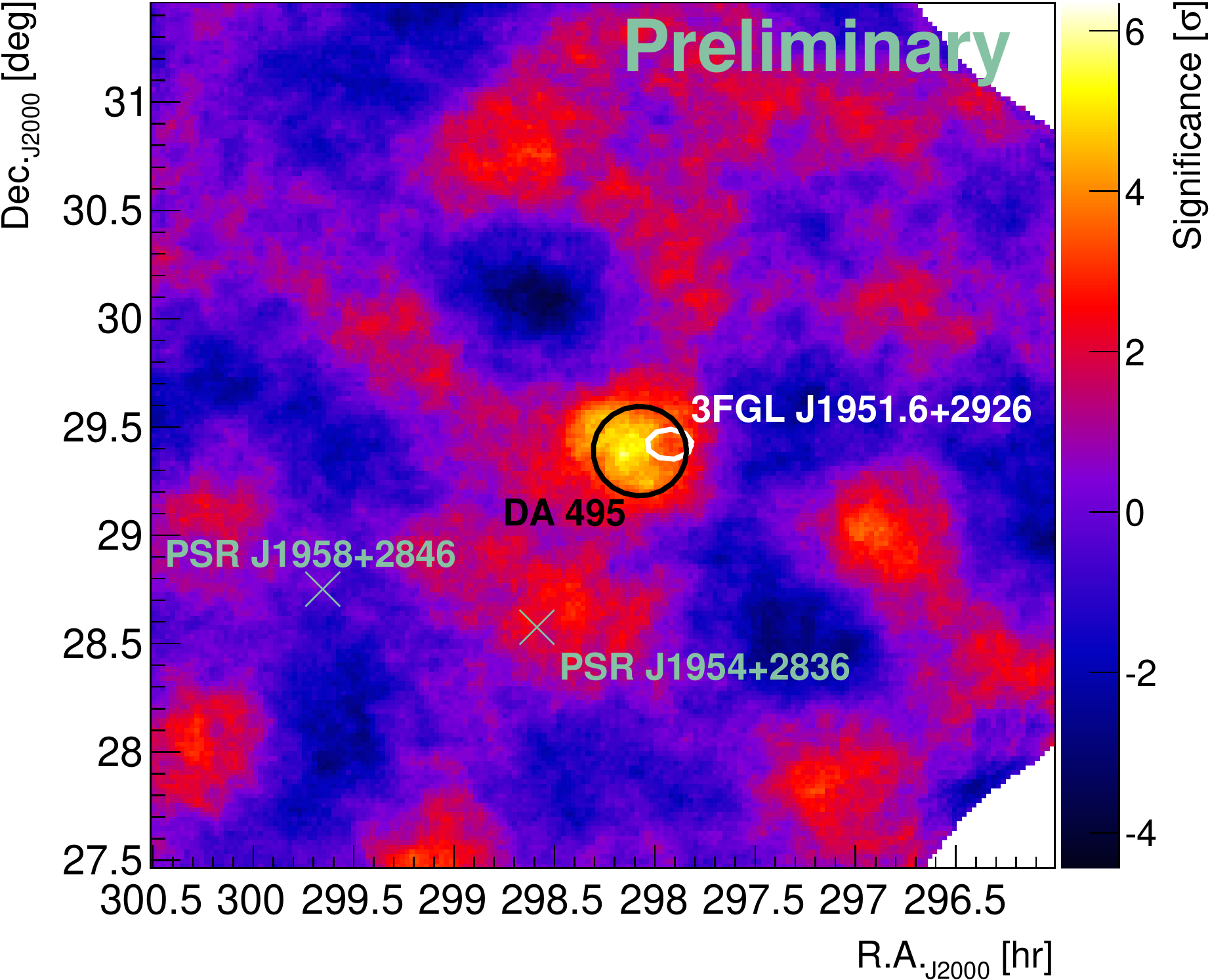}}
  \caption{VERITAS significance sky maps for two regions containing
    HAWC sources from the preliminary 1-year catalog. On the left is
    the region around the VERITAS source VER~J1930+188, likely
    associated with the PWN in SNR~G54.1+0.3, in which three HAWC
    sources have been identified. White crosses indicate the two HAWC
    sources without a VERITAS counterpart. White circles indicate
    regions excluded from the background estimation. On the right is
    the region around another HAWC source, 2HWC~J1953+294. The
    brightest VERITAS emission overlaps with the PWN DA~495.}
  \label{HAWC}
\end{figure}

VERITAS has 27 hours of archival data on this region, taken in
standard ``wobble'' mode with $0.5^{\circ}$ offset around
G54.1+0.3. All of these data were recorded prior to 2009 (before any
of the upgrades to VERITAS were implemented), and half are with only 3
of the 4 telescopes in operation. We have supplemented these data with
an additional 12 hours of recent observations (June 2016), targeting a
point roughly equidistant from the 3 HAWC sources. The resulting
skymap, for the complete dataset, is shown on the left in
Figure~\ref{HAWC}. VER~J1930+188 is re-detected, as we would expect,
but there is no evidence for emission from either of the new HAWC
sources. Studies are ongoing, but the non-detection may be explained
by a combination of effects: for example, the limited sensitivity of
the archival data for sources with large offsets, and/or the influence
of source characteristics such as spatial extension and hard spectra.

Dedicated VERITAS observations have also been performed this summer of
another of the preliminary HAWC source candidates, 2HWC~J1953+294. The
VERITAS dataset in this case totals 37 hours, and reveals a
statistically significant gamma-ray source located within the HAWC
source contours (Figure~\ref{HAWC}, right). The emission is
coincident with the PWN DA~495 (G65.7+1.2) and its central object,
WGA~J1952.2+2925. Karpova et al. \cite{2015MNRAS.453.2241K} have
argued that, despite the lack of observed pulsations, the Fermi-LAT
source 3FGL~J1951.6+2926 is likely the counterpart of the pulsar in
this system, and not of the PWN. The addition of VHE measurements from
HAWC and VERITAS to the broadband SED will soon help to determine the
nature of the high energy emission from this interesting object.

\section{SUMMARY}
VERITAS continues to operate smoothly, and the instrument is currently
in its most sensitive configuration to-date. Ongoing analysis
developments promise further incremental improvements in the coming
years, while new scientific opportunities are presented by the
introduction of HAWC, IceCube, and other facilities. Operational
funding for VERITAS is secure, and the collaboration plans to continue
observing with the array until at least 2019, with a 10-year
anniversary workshop planned for 2017. The construction of the
prototype mid-size Schwarzchild-Couder Telescope \cite{pSCThere} at
the FLWO provides a firm link to CTA, and to the future of ground-based
gamma-ray astronomy.


\section{ACKNOWLEDGEMENTS}

This research is supported by grants from the U.S. Department of
Energy Office of Science, the U.S. National Science Foundation and the
Smithsonian Institution, and by NSERC in Canada. We acknowledge the
excellent work of the technical support staff at the Fred Lawrence
Whipple Observatory and at the collaborating institutions in the
construction and operation of the instrument. The VERITAS
Collaboration is grateful to Trevor Weekes for his seminal
contributions and leadership in the field of VHE gamma-ray
astrophysics, which made this study possible.


\end{document}